\documentclass[conference]{IEEEtran}

\usepackage{algorithm,algpseudocode}
\usepackage{caption}
\usepackage[dvips]{graphicx}
\usepackage{amssymb}
\usepackage{multirow}
\usepackage{colortbl}
\usepackage{array, tabularx}
\usepackage{subfigure}
\usepackage{amsthm}
\usepackage{amsmath}
\usepackage{setspace}
\usepackage{soul}
\usepackage{bm}

\normalsize

\ifCLASSINFOpdf
\else
\fi

\IEEEoverridecommandlockouts
\begin{document}
%

\title{Secure Clustered Distributed Storage\\ Against Eavesdroppers}
%
%
%

\author{\IEEEauthorblockN{Beongjun Choi, Jy-yong Sohn, Sung Whan Yoon, and Jaekyun Moon}
\IEEEauthorblockA{School of Electrical Engineering\\
Korea Advanced Institute of Science and Technology\\
Daejeon, 34141, Republic of Korea\\
Email: bbzang10@kaist.ac.kr, jysohn1108@kaist.ac.kr, shyoon8@kaist.ac.kr, jmoon@kaist.edu}
\thanks{This work was supported by the National Research Foundation of Korea under grant no. 2016R1A2B4011298 and ICT R\&D program of MSIP/IITP. [2016-0-00563, Research on Adaptive Machine Learning Technology Development for Intelligent Autonomous Digital Companion]}}

\maketitle

\setcounter{page}{1}

\begin{abstract}
This paper considers the security issue of practical distributed storage systems (DSSs) which consist of multiple clusters of storage nodes.
Noticing that actual storage nodes constituting a DSS are distributed in multiple clusters, two novel eavesdropper models - the node-restricted model and the cluster-restricted model - are suggested which reflect the clustered nature of DSSs.
In the node-restricted model, an eavesdropper cannot access the individual nodes, but can eavesdrop incoming/outgoing data for $L_c$ compromised clusters.
In the cluster-restricted model, an eavesdropper can access a total of $l$ individual nodes but 
the number of accessible clusters is limited to $L_c$. 
We provide an upper bound on the securely storable data for each model, while a specific network coding scheme which achieves the upper bound is obtained for the node-restricted model, given some mild condition on the node storage size.
\end{abstract}


%
\IEEEpeerreviewmaketitle
\section{Introduction}
%
%
%
%






A distributed storage system (DSS) is a network of relatively inexpensive storage nodes to save data reliably over a long period of time.
A good example is the large data centers which use storage nodes widely spread over the Internet \cite{ref:TotalRecall}, \cite{ref:Dhash}.
Since these storage nodes are made of commodity storage devices, they often fail to provide high reliability.
In order to support reliable storage, a DSS provisions the node repairing process in the case of node failure events.
If an active node fails, a newcomer node joins the system by downloading data from some of the remaining active nodes.

When a DSS is composed of insecure storage nodes, system security must be guaranteed in the presence of the threat of an intruder. 
Many researchers have studied ways to secure DSSs against possible intruders \cite{ref:Poor2013}-\cite{ref:Evemodel2011}.
Roughly, intruders can be classified into two types. One is active intruders who can possibly change stored data in the system, and the other is passive eavesdroppers who can only read data stored in the system.
The present paper focuses on the passive eavesdropper model.
An eavesdropper model for a DSS is typically designed as a node-based intruder,
which can read stored data and downloaded data (during the repair process) for certain compromised storage nodes \cite{ref:Security2011}, \cite{ref:Evemodel2011}.



In the real world, storage nodes in the data centers are divided into many clusters called the racks \cite{ref:Rack2}.
A node in the rack can only communicate with nodes in other racks via top-of-rack switches \cite{ref:NetworkTraffic}.
Considering a real world DSS, the existing eavesdropper models may not fully capture the behavior of real intruders.
In this paper, we introduce two eavesdropper models - `node-restricted' and `cluster-restricted' - which reflect the clustered nature of the DSS.

In the node-restricted model, $L_c$ clusters are compromised by eavesdroppers, who cannot access the individual storage nodes residing in the clusters. The eavesdroppers in this model can read the data transmitted during the node repair processes, both incoming and outgoing data for the compromised clusters.
This model applies to the scenarios where an eavesdropper can only read data transmission passing through the top-of-rack switches.
The cluster-restricted eavesdropper is an intruder who can access the individual nodes, while the number of accessible clusters is limited to $L_c$.
This model pertains to the scenarios where the clusters are dispersed widely, so that the eavesdropper can hardly access the entire clusters.
Although we mainly visualize the rack as a cluster in this paper, the application of the suggested models is not constrained to the multi-rack structure; the models can be used for general clustered DSS scenarios.



The main contribution of this paper is to design and analyze the two eavesdropper models well-suited to clustered distributed storage systems.
We also provide upper bounds on securely storable data for the suggested eavesdropper models. 
Furthermore, an explicit coding scheme to achieve the theoretical limit of secrecy capacity is suggested in the node-restricted model.

This paper is organized as follows. Section \ref{section:2} reviews the dynamic structure of the distributed storage system and secrecy capacity.
In Sections \ref{section:3} and \ref{section:4}, we suggest and analyze two eavesdropper models.
Finally, conclusions are drawn in Section \ref{section:5}.

\section{Backgrounds}\label{section:2}
\subsection{Distributed Storage Systems}
A DSS is a network of storage nodes used to store data. 
Suppose a source has an incompressible file $\mathcal{F}$ that needs to be stored in the system. 
The source encodes a file to ensure reliability and distributes to a certain number $n$ of storage nodes, each of which has a storage size $\alpha$.
The system is designed in a way that any users (also called data collectors) can read the original file $\mathcal{F}$ by contacting any choice of   $k<n$ storage nodes.

However, individual storage nodes in a DSS frequently fail to provide reliable information to the data collector, degrading the reliability of the system.
In order to store data for a long period of time, the node repair process is required to keep $n$ active storage nodes.
When an active storage node fails, an inactive node in the system receives information from $d<n$ active nodes to newly join the system.
Let the total amount of data that the newcomer node receives to join the system be $\gamma$ (total repair bandwidth), and the contribution of each active node be $\beta=\gamma/d$.
We denote a DSS with parameters $n, k$ as $\mathcal{D}(n,k)$.





\subsection{Information Flow Graph}
A DSS can be represented as an information flow graph in the sense that it can describe a flow of information between storage nodes \cite{ref:Dimakis}.
The graph consists of three types of nodes: a source node, data collector nodes and storage nodes.
A source node stores the encoded file into multiple storage nodes.
The data collector nodes receive the data from storage nodes to reconstruct the original file.
Each storage node $v_{i}$ is divided into the input storage node $v_{i}^{in}$ and the output storage node $v_{i}^{out}$.
They are connected by a directed edge with capacity $\alpha$ to represent the storage size of a node.

\begin{figure}[!t]
    \centering
    \includegraphics[height=50mm]{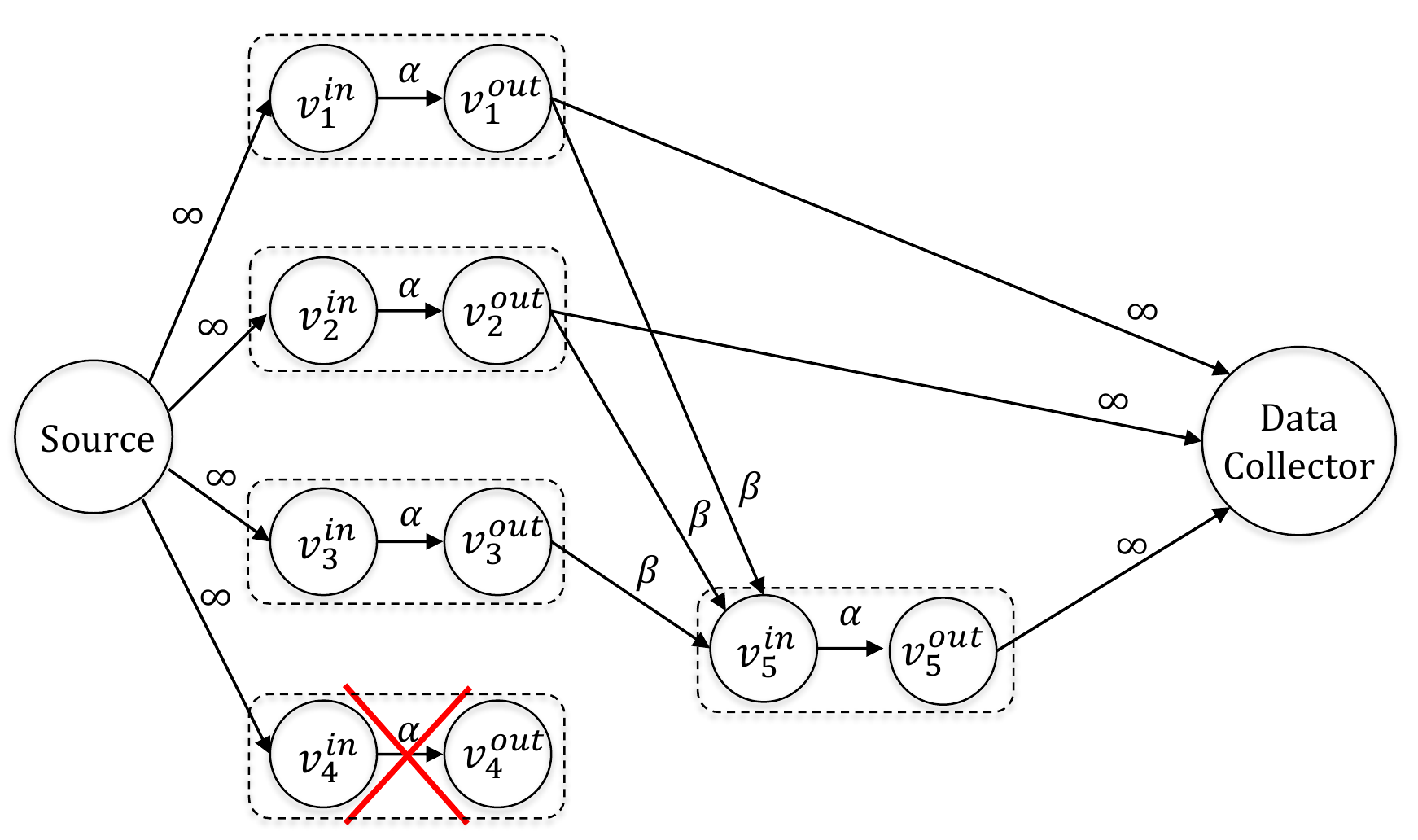}
    \caption{Information flow graph for $\mathcal{D}(n = 4, k = 3)$ with $d=3$}
    \label{Fig:Flowgraph}
\end{figure}

In Fig. \ref{Fig:Flowgraph}, the encoded file gets stored into $n$ storage nodes. An edge with infinite capacity connects the source node and each storage node $v_{i}$ for $i=1,\dots,n$.
When a node fails, a newcomer node joins the system by receiving information $\beta$ from $d$ active helper nodes. This is illustrated as $d$ edges with capacity $\beta$.
A data collector node connects to $k$ active nodes by edges with infinite capacity to reconstruct the original file.

A cut between a source node and a data collector is defined as a set $C$ of edges which satisfies the following: every directed path from the source to the data collector must pass at least one edge in $C$. 
The minimum cut $C^*$ is defined as a cut $C$ separating the source and the data collector with the smallest sum of edge capacities.

Due to the dynamic nature of the failure events and the repair processes in a DSS, the number of possible flow graphs of a DSS can be infinite.
The authors of \cite{ref:Dimakis} has shown that if a minimum cut value separating a source node and a data collector node is larger than or equal to certain amount $\mathcal{M}$ (for all possible flow graphs), there exists a linear network coding scheme to store data $\mathcal{M}$ to a DSS.


\subsection{Eavesdropper Model and Secrecy Capacity}
In a DSS, an eavesdropper is typically viewed as a passive intruder who can access the individual storage nodes and read the data from the compromised nodes without altering them.
It is generally assumed that an eavesdropper has a complete knowledge of the storage system and the repair scheme in the case of node failure.
The eavesdropper models for a DSS are suggested in \cite{ref:Security2011} and \cite{ref:Evemodel2011}, where both models assume that the intruder
can access the individual nodes.
The eavesdropper in \cite{ref:Security2011} is characterized by a parameter $l<k$ which represents the power of the intruder. She is able to read the downloaded data (during the node repair process) as well as stored data from $l$ compromised nodes.
Therefore, an eavesdropper may choose to intrude the currently active nodes or temporary inactive nodes which will read the message during the node repair process.
On the other hand, the authors of \cite{ref:Evemodel2011} distinguished the intrudable storage nodes into two types: one can read the stored data only, while the other can read both the stored data and the downloaded data.
We basically adopt the eavesdropper model suggested in \cite{ref:Security2011} for Section \ref{section:4}.

The maximum amount of information that can be stored with perfect secrecy against the eavesdropper
is defined as the secrecy capacity $C_{s}(\alpha,\gamma)$.
In \cite{ref:Security2011}, an upper bound of the secrecy capacity is derived as
\begin{equation} \label{Eq:Csref}
C_{s}(\alpha,\gamma) \leq \sum_{i=l+1}^{k}\min\{ (d-i+1)\beta, \alpha \}.
\end{equation}
This upper bound is calculated as the minimum cut value of the information flow graph excluding the compromised links.
Although calculating the exact value of the secrecy capacity for an arbitrary DSS still remains as an open problem, the  authors of \cite{ref:RSKR} provided an explicit coding scheme to achieve the upper bound when the storage size $\alpha$ is sufficiently large.

\section{Node-Restricted Eavesdropper Model}\label{section:3}
To begin with, a model for clustered DSSs is introduced with appropriate parameters. 
A clustered DSS is defined as a DSS consisting of multiple clusters where a single cluster has multiple storage nodes. 
We assume that every cluster has the equal number of active storage nodes.
We use the parameters $L$ and $n_{I}=n/L$ to represent the number of total clusters and the number of active storage nodes in each cluster, respectively. 
In a node repair process, increasing the number $d$ of helper nodes  is always beneficial in terms of the amount of securely stored data in the system \cite{ref:Security2011}. 
Thus we assume that $d$ has the maximum possible value $n-1$ throughout the paper.
Also, we assume that the failed node and the corresponding newcomer node reside in the same cluster to keep the consistent number of nodes across clusters.



In this section, we suggest a \textit{node-restricted eavesdropper model} wherein the intruder cannot access the individual nodes but read data transmission between different clusters. 
In Hadoop distributed file systems \cite{ref:Rack2}, \cite{ref:Hadoop} as well as other large scale data centers, storage nodes are dispersed to multiple racks to easily deal with a large number of storage nodes.
	The nodes within the same rack are connected via a top-of-rack switch to enable communication of nodes in different racks.
	Communication between two nodes residing in the same rack, however, does not need to go through a switch.
	It is reasonable to assume that an eavesdropper compromises the top-of-rack switches which are the cores of the cross-rack communication.
	In this case, the eavesdropper cannot read individual nodes but may access the information passing through the switches.
	To this end, we develop a node-restricted eavesdropper model with parameter $L_c$, the number of compromised clusters (top-of-rack switches).




\begin{figure}[!t]
    \centering
    \includegraphics[height=38mm]{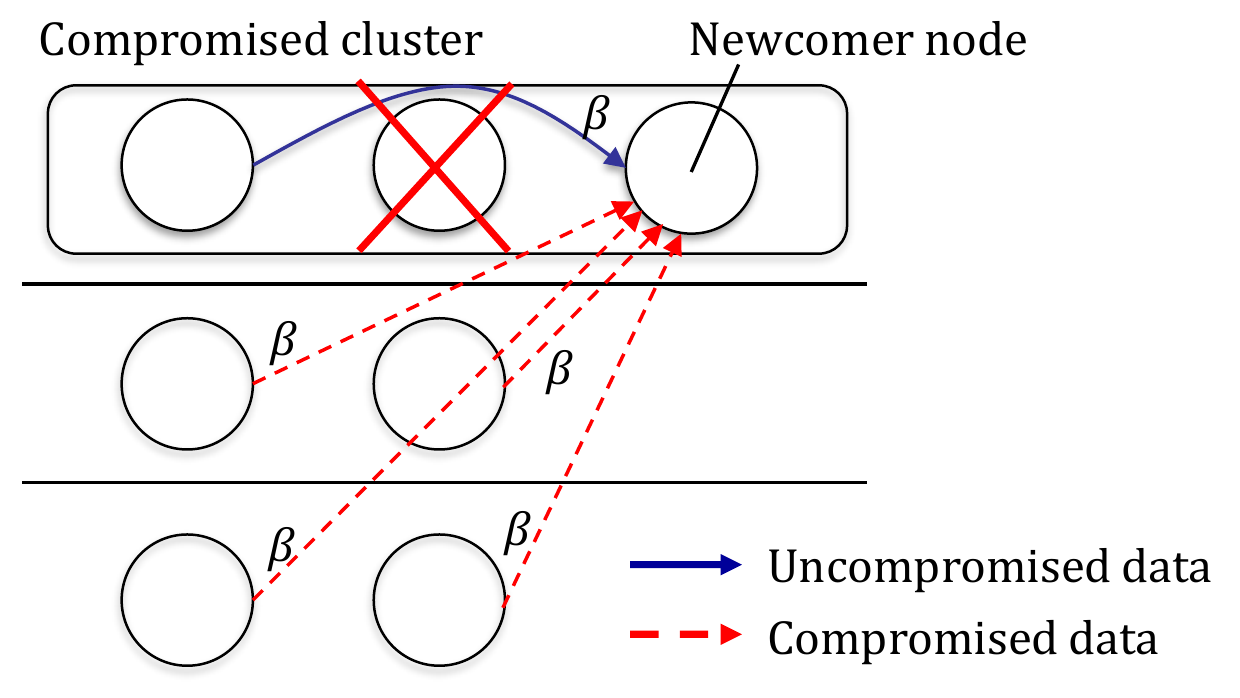}
    \caption{The node-restricted eavesdropper model. The eavesdropper can read incoming/outgoing data for compromised clusters but cannot access the individual nodes.}
    \label{Fig:node-restricted}
\end{figure}



\newtheorem{theorem}{Theorem}
\newtheorem{lemma}{Lemma}[theorem]
\newtheorem{corollary}{Corollary}
\newenvironment{definition}[1][Definition]{\begin{trivlist}
\item[\hskip \labelsep \normalfont #1]}{\end{trivlist}}

\begin{theorem} \label{Theorem:Theorem1}
(Node-restricted eavesdropper : upper bound of secrecy capacity with maximum helper nodes) For a clustered distributed storage system $\mathcal{D}(n,k)$ with $L_c\leq L$ compromised clusters, the secrecy capacity $C_{s}(\alpha,\gamma)$ is upper bounded by 
\begin{equation}\label{equation:3.1}
C_{s}(\alpha,\gamma) \leq 
\begin{cases}
L_{c} \sum_{i=1}^{n_{I}}\min\{(n_{I}-i)\beta,\alpha\} \\ + \sum_{i=n_{I}L_{c}+1}^{k}\min\{(n-i)\beta,\alpha\}, & n_{I}L_{c}<k \\\\
\lfloor k/n_{I} \rfloor \sum_{i=1}^{n_{I}}\min\{(n_{I}-i)\beta,\alpha\} \\ +\sum_{i=1}^{mod(k,n_{I})}\min\{(n_{I}-i)\beta,\alpha\}, & \mbox{otherwise}
\end{cases}
\end{equation}
where $\beta=\gamma/d$.
\end{theorem}
\begin{proof}
%
%
%
%
Case 1) $n_{I}L_{c}<k$.

Let the encoded file be distributed into $n$ storage nodes $v_{1},v_{2},\dots,v_{n}$. 
Let $V=\{v_{1},\dots,v_{n}\}$.
Assume that nodes $v_{1+(i-1)n_{I}},\dots,v_{in_{I}}$ reside in the $i$th compromised cluster for $1\leq i\leq L_{c}$.
Assume nodes $v_{1},v_{2},\dots,v_{k}$ fail consecutively and then are replaced by $v_{n+1},v_{n+2},\dots,v_{n+k}$ via the successive node repairing processes as shown in Fig. \ref{Fig:GKproof}.

For $1 \leq i \leq k$, the edges coming into $v_{n+i}^{in}$ from $v \in V$ can be divided into two types: 
the solid edges connecting the nodes in the same cluster and the dotted edges connecting the nodes in different clusters.
The node-restricted eavesdropper can only read cross-cluster communication.
Therefore, the solid edges represent information which is not read by the eavesdropper.


Assume that the data collector reconstructs the original file by contacting $k$ nodes, $v_{n+1},v_{n+2},\dots,v_{n+k}$.
Let $C_i$ be the maximum amount of secure information that can be transmitted from the node $v_{n+i}$ to the data collector.
Then, the amount of information securely stored in the system is upper bounded by the entropy $H(C_{1},C_{2},\dots,C_{k})=\sum_{i=1}^{k}H(C_{i}|C_{1},\dots,C_{i-1})$.
Note that for every $i \in \{1, \dots, k\}$, $H(C_{i}|C_{1},\dots,C_{i-1})$ can be calculated as the minimum value associated with the incoming solid edges to $v_{n+i}$ as well as storage size of node $v_{n+i}$ (from Fig. \ref{Fig:GKproof}). Thus, we complete the proof by specifying each $H(C_{i}|C_{1},\dots,C_{i-1})$ value.

\begin{figure}[!t]
	\centering
	\includegraphics[height=52mm]{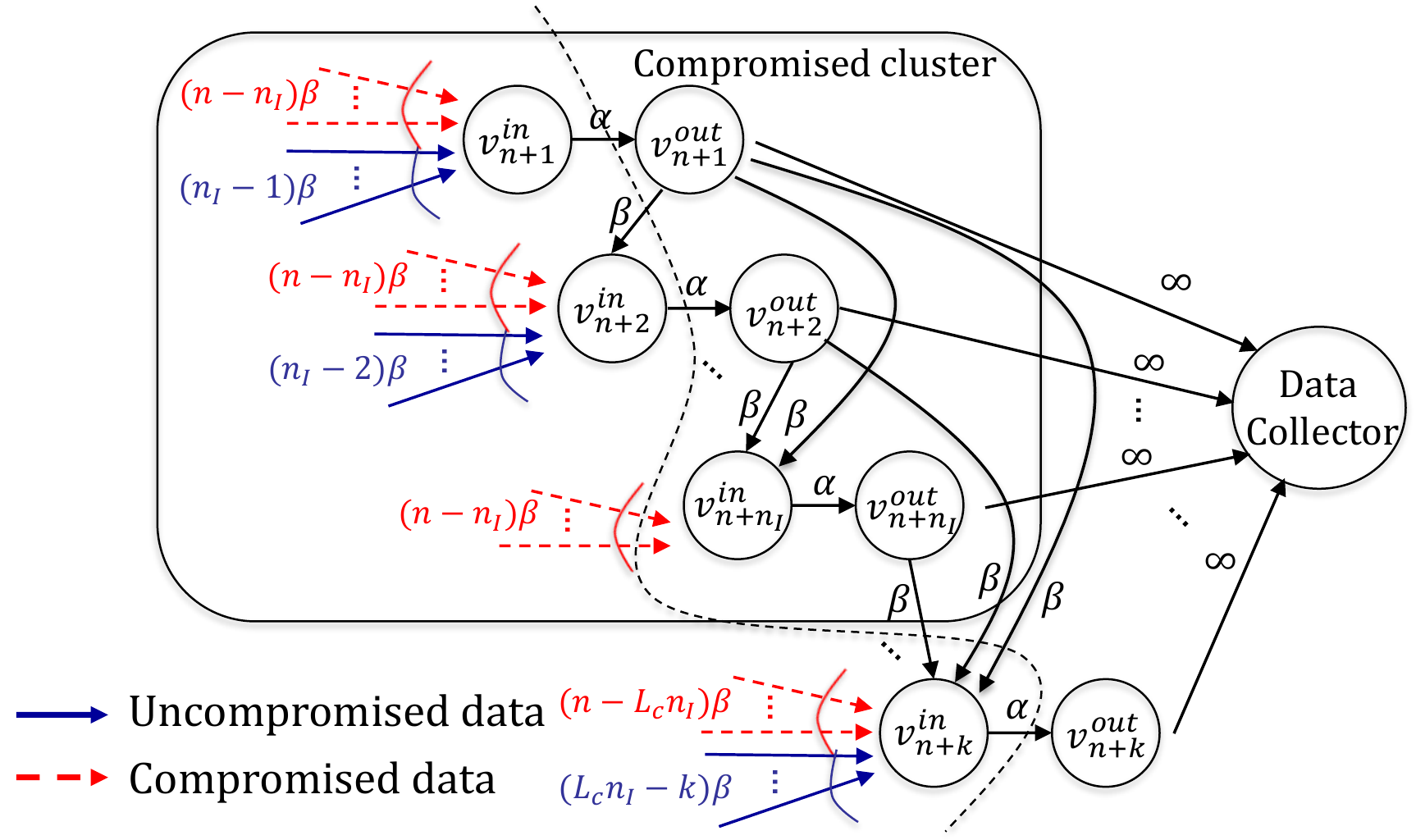}
	\caption{Information flow graph used in proving Theorem 1.}
	\label{Fig:GKproof}
\end{figure}

Case 2) $n_{I}L_{c} \geq k$.

The upper bound of secrecy capacity can be proved in a similar way by letting the nodes $v_{1+(i-1)n_{I}},\dots,v_{in_{I}}$  be in the $i$th compromised cluster ( $1\leq i\leq \lceil k/{n_{I}} \rceil$).

\end{proof}

\begin{figure}[!t]
    \centering
    \includegraphics[height=67mm]{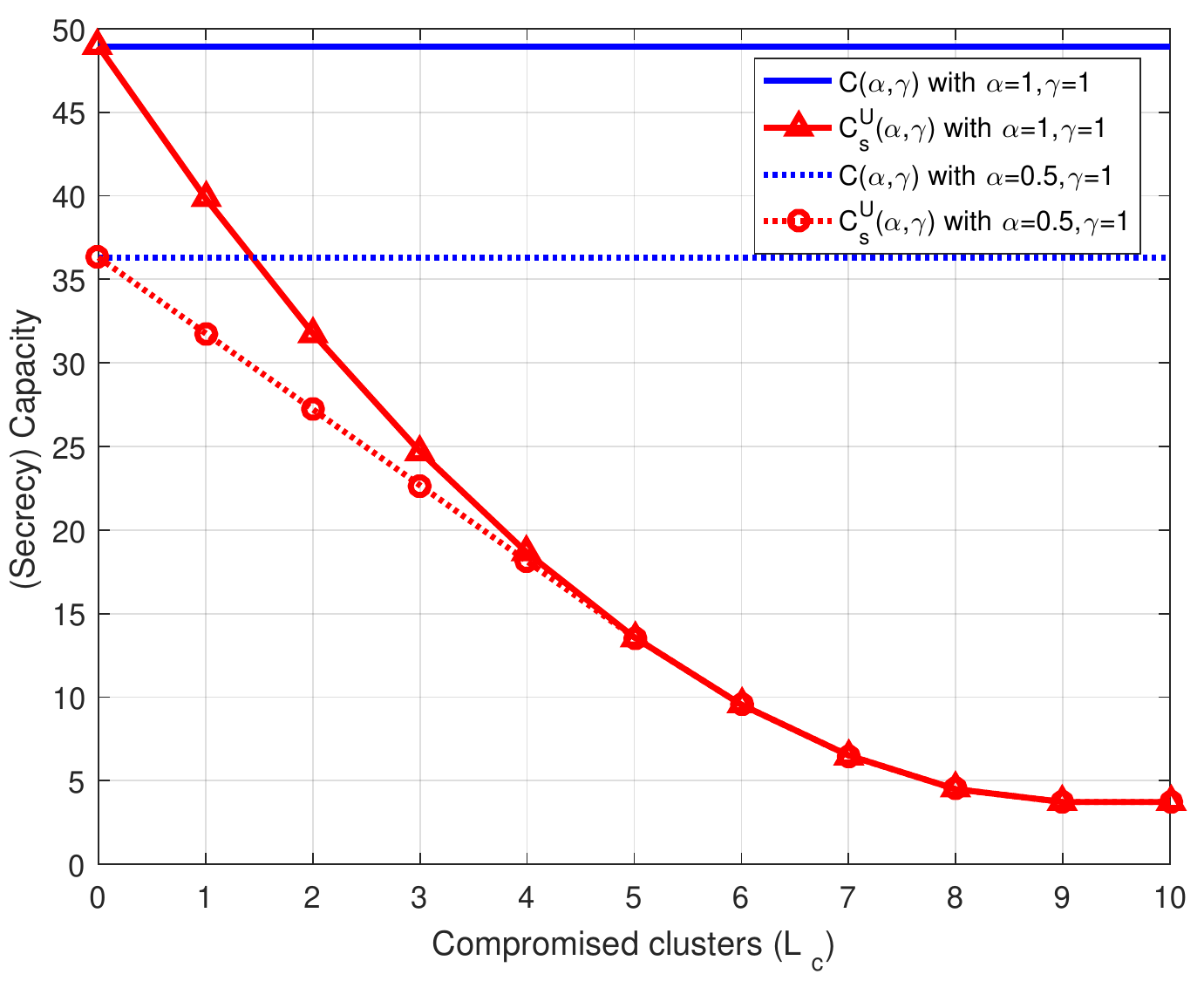}
    \caption{$C{(\alpha,\gamma)}$ and $C_{s}^{U}{(\alpha,\gamma)}$ (for a node-restricted model) of a clustered DSS $\mathcal{D}(n=100,k=85)$ with $L=10$ and $\gamma=1$}
    \label{Fig:node-restricted2}
\end{figure}

Fig. \ref{Fig:node-restricted2} shows the relationship between the number of compromised clusters and the upper bound of the secrecy capacity.
The capacity $C(\alpha,\gamma)=\sum_{i=1}^{k}\min\{(n-i)\beta,\alpha\}$ is the maximum amount of data that can be stored in a DSS without considering the eavesdropper \cite{ref:Dimakis}. The upper bound $C_{s}^{U}(\alpha,\gamma)$ denotes the RHS of inequality (\ref{equation:3.1}).

\begin{figure}[!t]
    \centering
    \includegraphics[height=38mm]{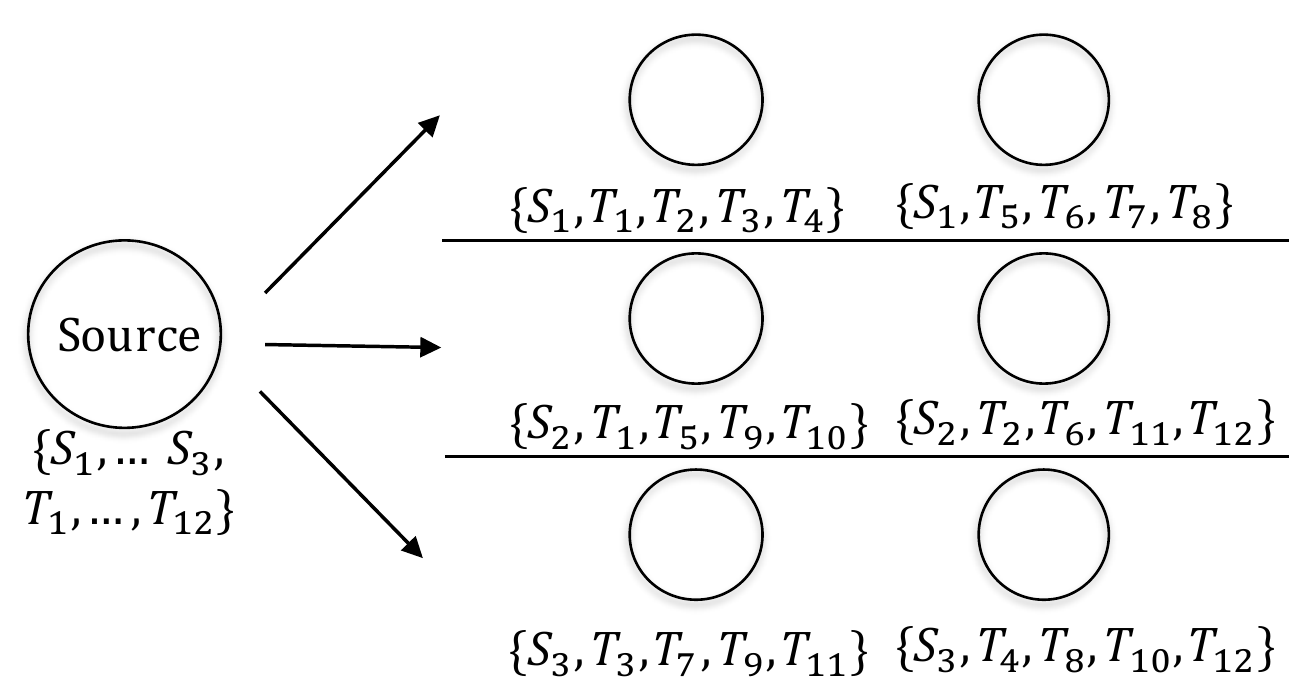}
    \caption{Example of RSKR network coding in a DSS $\mathcal{D}(n=6,k=5)$ with $d=5,L=3,$ $\alpha=5$ and $\gamma=5$ }
    \label{Fig:RSKR}
\end{figure}

It is shown that the upper bound derived in Theorem \ref{Theorem:Theorem1} can be achieved in the bandwidth-limited regime (when the size of storage nodes $\alpha$ is large enough) by an explicit coding scheme, called the RSKR \cite{ref:RSKR} repetition code.
The RSKR repetition code is a network coding scheme where each encoded symbol is stored on exactly two nodes while any choices of two storage nodes share one coded symbol (Fig. \ref{Fig:RSKR}).
This coding scheme also achieves the upper bound on secrecy capacity of the eavesdropper model in \cite{ref:Security2011}.


Here we provide an example of the RSKR repetition code which achieves the upper bound derived in Theorem \ref{Theorem:Theorem1}.
Consider a DSS $\mathcal{D}(n=6,k=5)$, with $d=5, \alpha=5,\beta=1,L=3$ and $L_{c}=1$ (Fig. \ref{Fig:RSKR}).
One can store $C(\alpha,\gamma) = 15$ symbols in the system, 
and securely store at most 
$C_{s}^{U}(\alpha,\gamma)=7$ symbols. 
Let the seven information symbols $I=\{I_{1},\dots,I_{7}\}$ be encoded to 15 symbols. For convenience, we classify 15 encoded symbols into 2 groups, $S=\{S_{1},\dots,S_{3}\}$ and $T=\{T_{1},\dots,T_{12}\}$. 
A source distributes 15 encoded symbols in the following manner.
Every element in the set $S$ is duplicated and stored in two nodes in a cluster, while different elements in $S$ are stored into distinct clusters.
Every element in the set $T$ is duplicated and stored in two nodes which reside in different clusters. The distributed 15 encoded symbols satisfy the RSKR property as in Fig. \ref{Fig:RSKR}.

Since the encoded symbols in the set $S$ is not transmitted across the cluster during the repair process, the node-restricted eavesdropper cannot access the encoded symbols in the set $S$ regardless of the choices of compromised clusters.
Therefore, three symbols in $S$ are stored in the system without the encoding process.
Notice that the eavesdropper with $L_{c}=1$ can read at most 8 encoded symbols in the set $T$ regardless of the choice of the compromised cluster.
This problem can be viewed as the wiretap channel II \cite{ref:wiretapII} with the parameter ($N=12$, $\mu=8$), where $N$ is the length of the encoded bits and $\mu$ is the number of eavesdropped bits by the intruder.
Thus, $N-\mu = 4$ information symbols can be securely encoded in $T$ with perfect secrecy. 
In summary, three (uncoded) information symbols in $S$ and four information symbols which are encoded in $T$ are securely storable. 
The total number of secure information symbols are equal to $7$, which coincides with the upper bound of Theorem \ref{Theorem:Theorem1}.

We can expand the example for general parameters $n, L$ and $L_{c}$. We may assume that $\beta =1$; generalization to arbitrary $\beta$ is easy by applying the same coding scheme parallelly.
For a distributed storage system $\mathcal{D}(n,k=n-1)$ with the $\alpha \geq d\beta$ condition (i.e., the bandwidth-limited regime), $C(\alpha,\gamma)=$ $ n\choose2$ and $C_{s}^{U}(\alpha,\gamma)$$=L_{c}$$n_{I}\choose2$ + $n-n_{I}L_{c}\choose 2$.
Suppose the RSKR coding scheme is applied. 
Let the set $S$ be the collection of symbols stored in two distinct nodes within the same cluster, and $T$ be collection of symbols stored in two nodes residing in different clusters.
Then, $\vert S \vert= $$L$$n_{I}\choose2$, $\vert T \vert= $$n\choose2$$-L$$n_{I}\choose2$.
The symbols in $S$ cannot be exposed to the eavesdropper, while at most  $\sum_{i=1}^{L_{c}}n_{I}(n-in_{I})$  symbols in $T$ may be exposed to the eavesdropper.
Therefore, the number of symbols securely stored in the system is $|S|+|T|-\sum_{i=1}^{L_{c}}n_{I}(n-in_{I})$.
With simple calculation, it is easy to verify that this value is equal to $C_{s}^{U}$. 
Thus, we conclude that the RSKR repetition code achieves the upper bound in Theorem \ref{Theorem:Theorem1} in the bandwidth-limited regime.


We expect another advantage of the RSKR repetition code based on its ability to make use of different data types.
Notice that the encoded symbols in the set $S$ can be systematically stored in the DSS. The data collector can reconstruct a symbol in $S$ by contacting a single node.
However, each symbol in the set $S$ is stored in a single specific cluster, so that the failure event of the cluster
causes irreversible data loss. 
The encoded symbols in the set $T$ are stored in a non-systematical way, so that the data collector should contact a sufficient number of nodes to collect the data in $T$.
On the other hand, the data encoded in $T$ is safe against any single cluster failure event.
Therefore, data which is frequently used but less important can be recommended to be stored in the set $S$, while important data which is not frequently used (such as private information) can be stored in the set $T$.



\section{Cluster-Restricted Eavesdropper Model}\label{section:4}

\begin{figure}[!t]
    \centering
    \includegraphics[height=30mm]{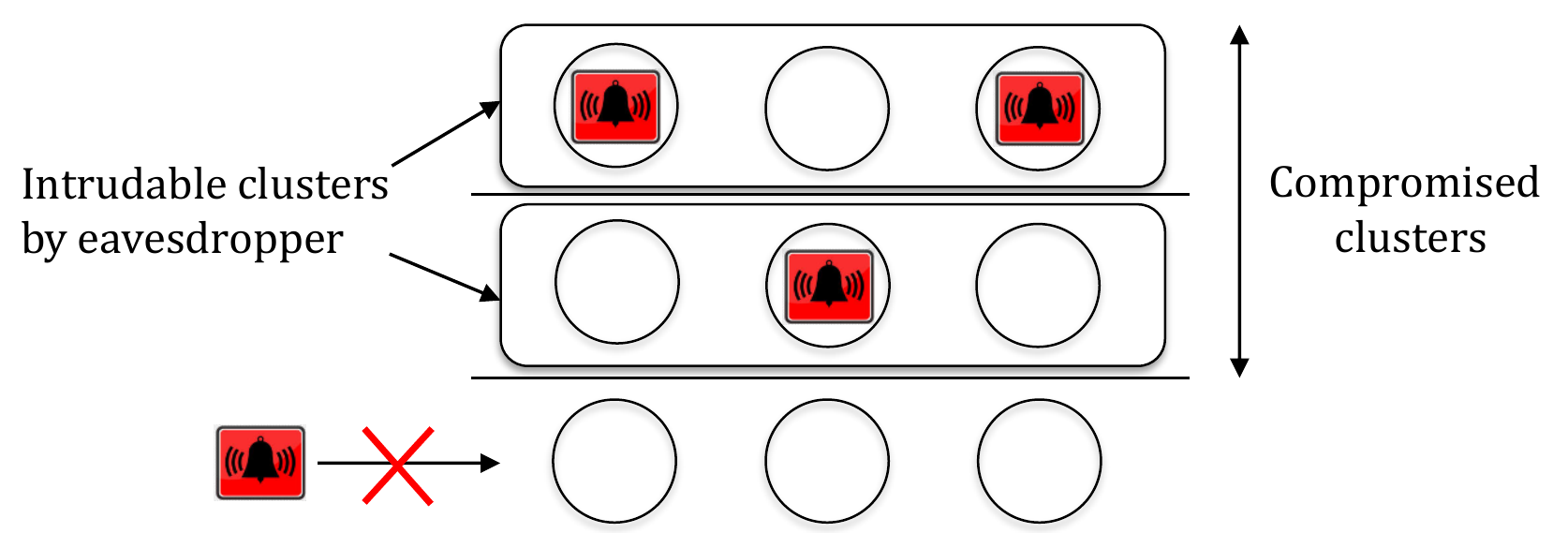}
    \caption{The cluster-restricted eavesdropper model. The eavesdropper can access individual nodes, but only a limited number of clusters.}
    \label{Fig:Intrudable}
\end{figure}

We introduce a \textit{cluster-restricted eavesdropper model} and provide an upper bound on the secrecy capacity of the suggested model.
As discussed, this type of eavesdropper can access individual $l \leq k$ storage nodes in the system;
however, the number of clusters that she can access is limited by $L_c \leq L$
(Fig. \ref{Fig:Intrudable}).
Here, we basically adopt a node eavesdropper model suggested in \cite{ref:Security2011}.
In other words, the suggested model with constraint $L_c = L$ reduces to the eavesdropper model in \cite{ref:Security2011}.
Trivially, inequality $l\leq n_{I}L_{c}$ is always satisfied.

\subsection{Symmetric Repair Model} \label{Section:symmetric}
The symmetric repair model is a node repair model where every newcomer node receives the equal amount of information $\beta$ from $d$ helper nodes during the node repair process.
The secrecy capacity of the cluster-restricted model with eavesdropper ($l,L_{c}$) is upper bounded as follows.
\begin{equation}\label{equation:Csbalanced}
C_{s}(\alpha,\gamma) \leq \sum_{i=l+1}^{k}\min\{ (n-i)\beta, \alpha \}
\end{equation}
Let the RHS of inequality (\ref{equation:Csbalanced}) be denoted as $C_{s}^{U}(\alpha,\gamma)$.
Notice that the upper bound $C_{s}^{U}(\alpha,\gamma)$ is consistent with the result of \cite{ref:Security2011}, irrespective of $L_{c}$. This result is due to the fact that every choice of compromised nodes in a clustered DSS equally affects the amount of securely storable data in the symmetric repair model.

\subsection{Asymmetric Repair Model}
Recently, a new repair model suitable for a clustered DSS was studied in \cite{ref:Sohn}.
The main idea of the suggested repair model is to distinguish the repair bandwidth $\beta$, depending on the relative location of the failed node and the helper node.
If the newcomer node and the helper node are located in the same cluster, a helper node transmits $\beta_{I}$ information.
Otherwise, a helper node transmits $\beta_{c}$ information to the newcomer node.
Considering a typical scenario of having larger intra-cluster communication bandwidth compared to the cross-cluster bandwidth, it was assumed that $\beta_{I}\geq \beta_{c}$.
This repair model can be viewed as a generalized version of the symmetric model in \cite{ref:Dimakis}.
Under this setting,
the total repair bandwidth of the asymmetric repair model is expressed as $\gamma = \gamma_{I} + \gamma_{c}$, where $\gamma_{I}$ and $\gamma_{c}$ are overall intra-repair and cross-repair bandwidths.



\begin{theorem} \label{Theorem:Theorem2}
(Cluster-restricted eavesdropper: upper bound of secrecy capacity in asymmetric repair model with maximum helper nodes) For a clustered distributed storage system $\mathcal{D}(n,k)$ with $l$ compromised nodes and  $L_c$ compromised clusters, the secrecy capacity $C_{s}(\alpha,\gamma_{I},\gamma_{c})$ is upper bounded by
\begin{equation}\label{equation:4.2}
C_{s}(\alpha,\gamma_{I},\gamma_{c}) \leq \sum_{i=1}^{n_I} \sum_{j=f(i)+1}^{g(i)} \min \{ x(i)\gamma_I + y^{*}(i,j) \gamma_c,\alpha\}
\end{equation}
where
\begin{align} \nonumber
f(i) &=
\begin{cases}
f_{1}(i), & L_{c} \leq \lfloor k/n_{I} \rfloor \\
f_{2}(i), & L_{c} >\lfloor k/n_{I} \rfloor, l \leq mod(k,n_{I}) (\lfloor k/n_{I} \rfloor +1) \\
f_{3}(i), & otherwise
\end{cases}\nonumber\\
\end{align}
\begin{align}
f_{1}(i)&=
\begin{cases}
L_{c}, &  i \leq \lfloor l/L_{c} \rfloor \\
l-L_{c} \lfloor l/L_{c} \rfloor , & i = \lfloor l/L_{c} \rfloor +1\\
0, & otherwise
\end{cases}\nonumber\\
f_{2}(i)&=
\begin{cases}
g(i), &  i \leq \lfloor \frac{l}{\lfloor k/n_{I} \rfloor +1} \rfloor \\
l- \lfloor k/n_{I} \rfloor \lfloor \frac{l}{\lfloor k/n_{I} \rfloor+1} \rfloor , & i= \lfloor \frac{l}{\lfloor k/n_{I} \rfloor +1} \rfloor +1 \\
0, & otherwise
\end{cases}\nonumber\\
f_{3}(i)&=
\begin{cases}
g(i), &  i \leq \lfloor \frac{l'}{\lfloor k/n_{I} \rfloor } \rfloor + mod(k,n_{I})  \\
l' - \lfloor k/n_{I} \rfloor \lfloor \frac{l'}{\lfloor k/n_{I} \rfloor} \rfloor, & i = \lfloor \frac{l'}{\lfloor k/n_{I} \rfloor } \rfloor \nonumber\\
& \ \ \ \ \ + mod(k,n_{I}) +1  \\
0, & otherwise
\end{cases}
\end{align}

\begin{align*} \nonumber
g(i)&=
\begin{cases}
\lfloor \frac{k}{n_{I}} \rfloor + 1 , & i \leq mod(k,n_{I}) \\
\lfloor \frac{k}{n_I} \rfloor, & otherwise
\end{cases} \nonumber \\
x(i)& = \frac{n_I - i}{n_I - 1}
\nonumber\\
y^{*}(i,j) &= 1-\frac{(l-i)+(\sum_{m=1}^{i-1}{\{g(m)-f(m)\}}+j-f(i))}{n - n_I}
\nonumber\\
l'&= l-mod(k,n_I) (\lfloor k/n_{I} \rfloor + 1) 
\end{align*}
\end{theorem}

We denote the RHS of inequality (\ref{equation:4.2}) as $C_{s}^{U}(\alpha,\gamma_{I},\gamma_{c})$, an upper bound of the secrecy capacity of the system.
The result of Theorem \ref{Theorem:Theorem2} is derived by showing the following:
1) For any choice of $k$ distinct storage nodes, 2) any choice of $l$ compromised nodes satisfying $L_{c}$ constraints and 3) any failure and repair order of storage nodes, 
the minimum cut value of the information flow graph $G$ is greater than or equal to the RHS of inequality (\ref{equation:4.2}).

\begin{figure}[!t]
    \centering
    \includegraphics[height=45mm]{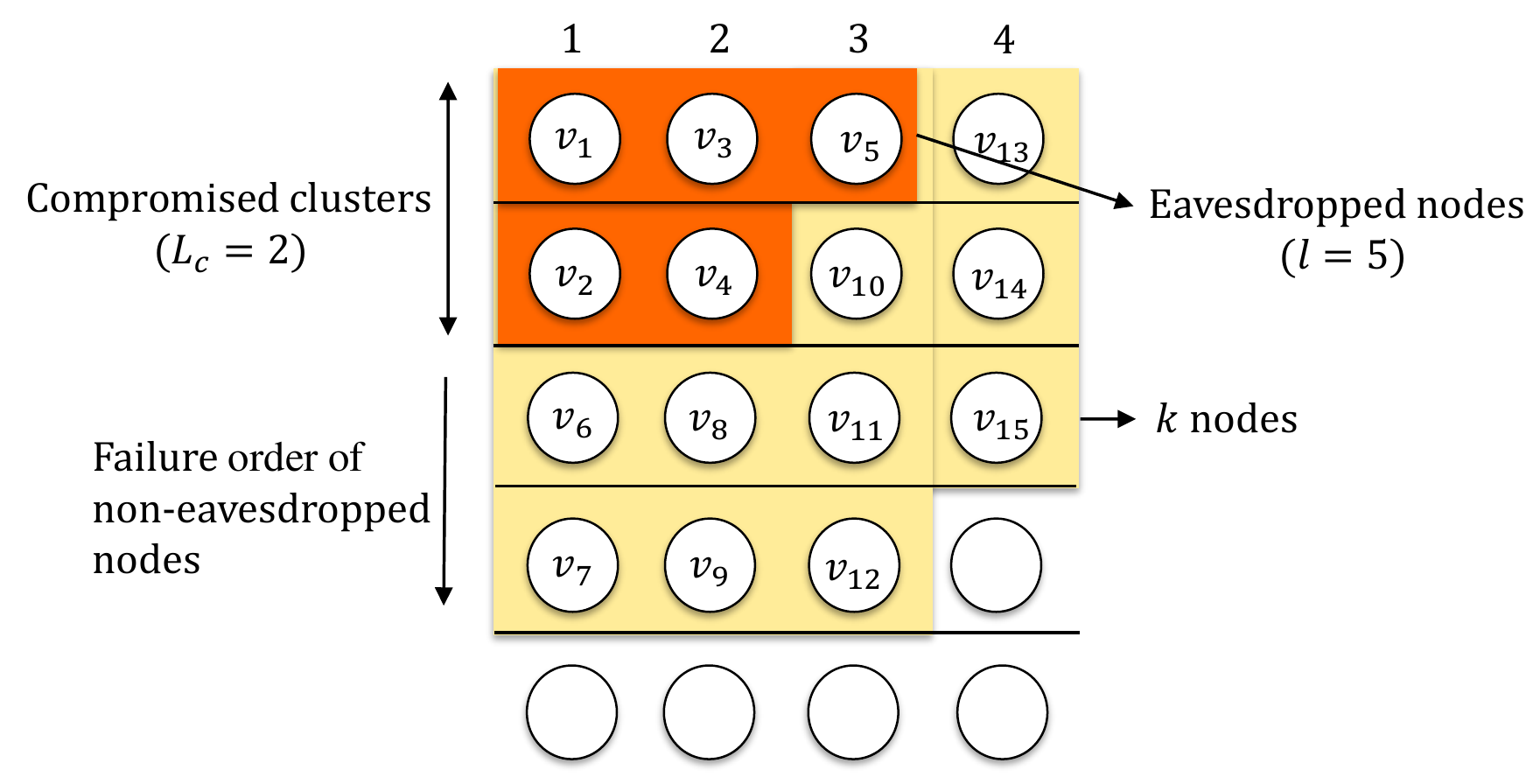}
    \caption{The way to design a flow graph $G^*$ to minimize the min-cut value in the asymmetric repair model for $\mathcal{D}(n=20,k=15)$ with $L=5$. The eavesdropper reads $l=5$ nodes inside the $L_{c}=2$ clusters. The failure order of $k=15$ nodes are specified inside the nodes.}
    \label{Fig:Failuresetorder}
\end{figure} 

The proof is omitted due to space limitation, but here we provide a sketch of the proof.
The full proof is available elsewhere.
$C_{s}^{U}(\alpha,\gamma_{I},\gamma_{c})$ is derived from the specific flow graph $G^*$ and obtained by minimum cut analysis.
We show that the minimum cut of possible flow graph (after removing compromised edges) is greater than or equal to the minimum cut of $G^*$.
Firstly, we figure out the optimal ordering of the failed nodes (similar to the vertical ordering in \cite{ref:Sohn}) which minimizes a minimum cut value with any choice of $k$ distinct storage nodes and any choice of $l$ compromised nodes satisfying the constraint $L_{c}$.
Secondly, we find the optimal choice of choosing $l$ eavesdropper nodes to minimize a minimum cut value with any choice of $k$ storage nodes and the optimal ordering method.
Finally, we find the optimal choice of choosing $k$ storage nodes to minimize a minimum cut value with the optimal choice of $l$ eavesdropper nodes and the optimal ordering method.

One way to design the flow graph $G^*$ satisfying the equality condition is given as follows. 
Let storage nodes $v_{1},\dots,v_{k}$ fail successively and be replaced by the nodes $v_{n+1},\dots,v_{n+k}$.
The data collector gathers information by connecting $k$ replaced storage nodes, $v_{n+1},\dots,v_{n+k}$.
The corresponding $k$ failed nodes $v_{1},\dots,v_{k}$ are selected such that the $k$ failed nodes belong to $\lceil k/n_{I} \rceil$ compromised clusters and $\lfloor k/n_{I} \rfloor$ clusters are full of failed nodes.
Then, choose $l$ eavesdropper nodes as $v_{n+1},\dots,v_{n+l}$. The corresponding failed nodes $v_{1},\dots,v_{l}$ are chosen to satisfy the number constraint $L_{c}$ and they are evenly spread across $L_{c}$ clusters.
Finally, select the failed nodes $v_{l+1},\dots,v_{k}$ successively such that in each of the $k-l-1$ steps taken, 
 the node in the cluster with a maximal number of remaining unfailed nodes is chosen.
This procedure is illustrated in Fig. \ref{Fig:Failuresetorder}.


\subsection{Discussions on Cluster-Restricted Eavesdropper}

\begin{figure}[!t]
    \centering
    \includegraphics[height=67mm]{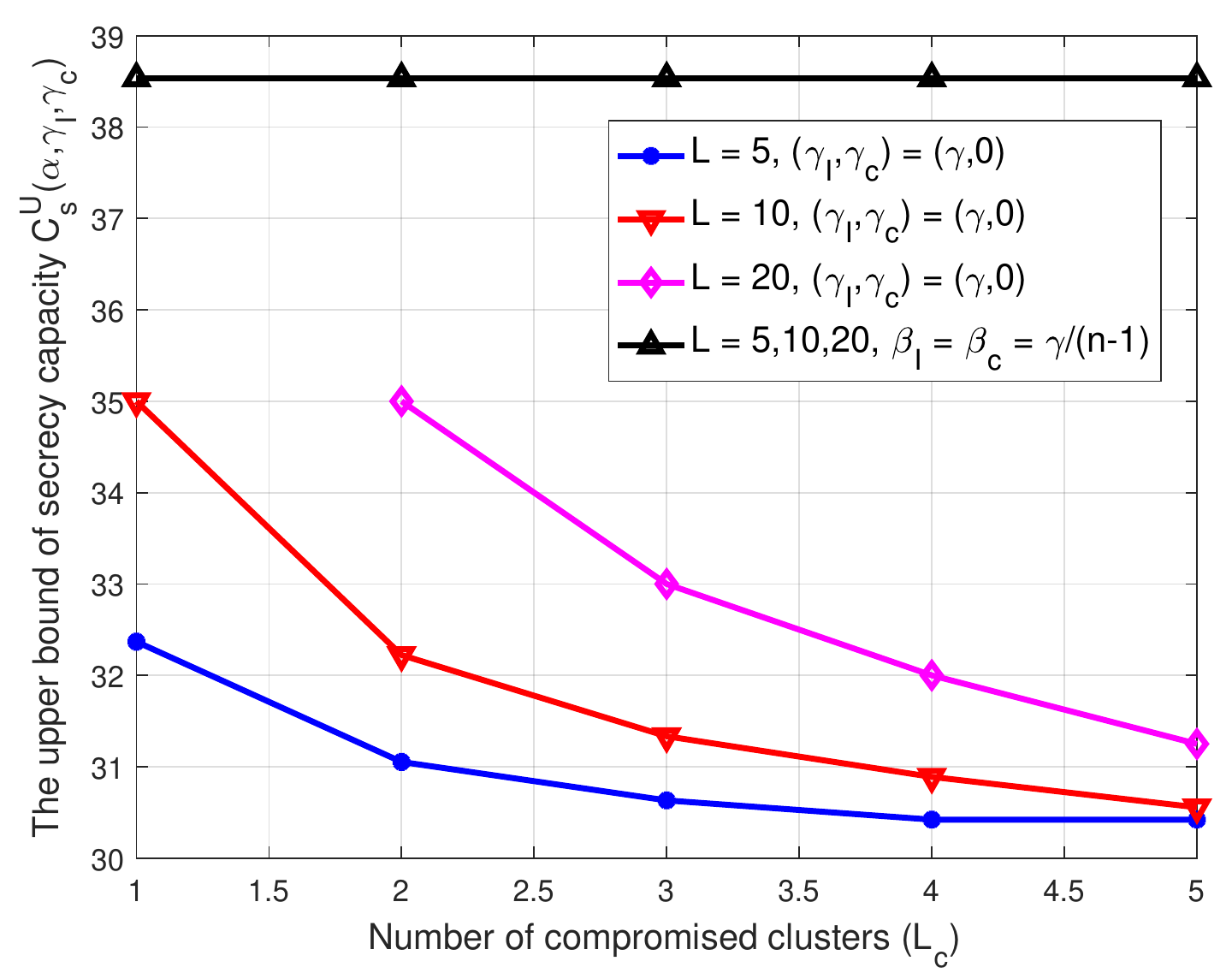}
    \caption{$C_{s}^{U}(\alpha,\gamma_{I},\gamma_{c})$ (for a cluster-restricted model) of a clustered DSS $\mathcal{D}(n=100,k=80)$ in the bandwidth-limited regime, $L=10$, $l=10$ and $\gamma = 1$}
    \label{Fig:Csconstrained}
\end{figure}

The upper bound of the secrecy capacity $C_{s}^{U}$ is a monotonic non-increasing function of $l$ and $L_{c}$.
The power of eavesdropper becomes stronger with larger $l$ and $L_{c}$, which results in small $C_{s}^{U}$.
Fig. \ref{Fig:Csconstrained} shows values of $C_{s}^{U}$ in the bandwidth-limited regime for a fixed number of compromised nodes $l$ with changing $L_{c}$ values.
Notice that $C_{s}^{U}$ in the symmetric repair model is independent of $L_{c}$ as confirmed in Section \ref{Section:symmetric}. 
However, $C_{s}^{U}$ is a decreasing function of $L_{c}$ in the asymmetric model. 

Since the cross-cluster repair bandwidth $\gamma_{c}$ is typically oversubscribed by a factor of $5-20$ \cite{ref:NetworkTraffic}, reducing $\gamma_{c}$ is quite demanding. 
It is shown in \cite{ref:Sohn} that $\gamma_{c}$ can be arbitrary reduced by spending more resources: the total repair bandwidth or the node storage size. 
We simulated under a $\gamma_{c}=0$ constraint in Fig. \ref{Fig:Csconstrained} to reduce the traffic passing the top-of-rack switches.
Notice that reducing the amount of cross-cluster communication traffic incurs a cost in terms of the amount of securely storable data in the system.




\section{Conclusion}\label{section:5}

We have introduced two eavesdropper models - the node-restricted model and the cluster-restricted model - which reflect the clustered nature of data centers in the real world. For each model, an upper bound of secrecy capacity is derived. Furthermore, an explicit coding scheme to achieve the upper bound is suggested for the node-restricted model.
Considering a realistic scenario where coded data blocks are allocated in multiple racks, the amount of securely storable data against an eavesdropper can be evaluated using our results. Analysis on the optimal dispersion strategy of coded blocks in multi-rack scenario as well as securely storable data against active adversary remains as an interesting future work.
\ifCLASSOPTIONcaptionsoff
  \newpage
\fi


\begin{thebibliography}{1}



\bibitem{ref:TotalRecall}
Bhagwan, Ranjita, et al. ``Total Recall: System Support for Automated Availability Management." \emph{NSDI}. Vol. 4. 2004.

\bibitem{ref:Dhash}
Dabek, Frank, et al. ``Designing a DHT for Low Latency and High Throughput." \emph{NSDI}. Vol. 4. 2004.

\bibitem{ref:Dimakis}
A. G. Dimakis, et al., ``Network Coding for Distributed Storage Systems," in \emph{IEEE Transactions on Information Theory}, vol. 56, no. 9, pp. 4539-4551, Sept. 2010.

\bibitem{ref:Poor2013}
T. Ernvall, et al., ``Capacity and Security of Heterogeneous Distributed Storage Systems," in \emph{IEEE Journal on Selected Areas in Communications}, vol. 31, no. 12, pp. 2701-2709, December 2013.

\bibitem{ref:Security2011}
S. Pawar, et al., ``Securing Dynamic Distributed Storage Systems Against Eavesdropping and Adversarial Attacks," in \emph{IEEE Transactions on Information Theory}, vol. 57, no. 10, pp. 6734-6753, Oct. 2011.

\bibitem{ref:Evemodel2011}
N. B. Shah, et al., ``Information-Theoretically Secure Regenerating Codes for Distributed Storage," \emph{GLOBECOM 2011}, 2011 IEEE, Houston, TX, USA, 2011, pp. 1-5.




\bibitem{ref:Rack2}
S. Muralidhar, et al. f4: Facebook's Warm Blob Storage System. In \emph{Proc. of USENIX OSDI}, 2014.

\bibitem{ref:NetworkTraffic}
Benson, et al., "Network traffic characteristics of data centers in the wild." \emph{Proceedings of the 10th ACM SIGCOMM conference on Internet measurement}. ACM, 2010.

\bibitem{ref:Hadoop}
Shvachko, Konstantin, et al. "The hadoop distributed file system." Mass storage systems and technologies (MSST), \emph{IEEE 26th symposium on. IEEE}, 2010.

\bibitem{ref:RSKR}
K. Rashmi, et al., ``Exact regenerating codes for distributed storage," in \emph{Proc. 47th Annu. Allerton Conf. Commun., Control, Comput.}, 2009.


\bibitem{ref:wiretapII}
L. H. Ozarow and A. D. Wyner, ``Wire-tap channel II," \emph{Bell Labs Tech. J.}, vol. 63, no. 10, pp. 2135-2157, dec 1984.

\bibitem{ref:Sohn}
J. Sohn, et al., ``Capacity of Clustered Distributed Storage", in arXiv:1610.04498v1, 2016.



\end{thebibliography}
\end{document}